# Hybrid-Vehcloud: An Obstacle Shadowing Approach for VANETs in Urban Environment

Anirudh Paranjothi[1], Mohammad S. Khan[2], Mohammed Atiquzzaman[1]
[1]School of Computer science, University of Oklahoma, – Norman, Oklahoma, USA.
[2]Department of Computing, East Tennessee State University, – Johnson City, Tennessee, USA.

*Abstract*— **Routing of messages in Vehicular Ad-hoc Networks (VANETs) is challenging due to obstacle shadowing regions with high vehicle densities, which leads to frequent disconnection problems and blocks radio wave propagation between vehicles. Previous researchers used multi-hop, vehicular cloud or roadside infrastructures to solve the routing issue among the vehicles, but they suffer from significant packet delays and frequent packet losses arising from obstacle shadowing. We proposed a vehicular cloud based hybrid technique called Hybrid-Vehcloud to disseminate messages in obstacle shadowing regions, and multi-hop technique to disseminate messages in non-obstacle shadowing regions. The novelty of our approach lies in the fact that our proposed technique dynamically adapts between obstacle shadowing and non-obstacle shadowing regions. Simulation based performance analysis of Hybrid-Vehcloud showed improved performance over Cloud-assisted Message Downlink Dissemination Scheme (CMDS), Cross-Layer Broadcast Protocol (CLBP) and Cloud-VANET schemes at high vehicle densities.**

## I. Introduction

Vehicular Ad-hoc Networks (VANETs) based on Vehicle-to-Vehicle (V2V) and Vehicle-to-Infrastructure communication (V2I) using Dedicated Short Range Communication (DSRC) can be used to disseminate messages, such as traffic jam, accident information, etc. between vehicles in a connected vehicular environment [1,2]. In VANET, messages transmitted among vehicles can be dropped due to intermittent connections resulting from obstacle shadowing from tall buildings, skyscrapers, etc. in regions, like Manhattan and other downtown areas with high vehicle densities.

Previous authors used either multi-hop (V2V), vehicular cloud or roadside infrastructures (V2I) such as RSUs to disseminate messages among the vehicles in non-obstacle shadowing regions. Carpenter *et al.* [3] and Nilsson *et al.* [4] proposed a model to validate the obstacle shadowing and presented the model to estimate the performance of VANET safety applications. However, it has limitations, such as high delay, and frequent loss of connectivity. Bi *et al.* [5] discussed multi-hop based protocols to disseminate messages between vehicles. However, it is not suitable for vehicle-dense obstacle shadowing regions like Manhattan environment. Syfullah *et al.* [6] and Liu *et al.* [7] discussed roadside unit for disseminating the critical messages to the nearby vehicles with the help of vehicular cloud network. However, this approach is not suitable for urban scenarios due to the long delays associated with the transmission of messages.

To overcome the obstacle shadowing problem in VANETs, we propose a hybrid technique called Hybrid-Vehcloud which adopts mobile gateways (such as busses) in the vehicular cloud for messages dissemination in obstacle shadowing regions, and multi-hop technique for regions with no obstacle shadowing.

In Hybrid-Vehcloud, mobile gateways, implemented using busses, are employed in obstacle shadowing regions due to the following reasons: 1) RSUs cannot be used as radio transmissions due to RSUs are severely affected by obstacles in vehicle-dense regions leading to frequent loss of packets, 2) antennas placed on buses can be placed higher than on cars, and thereby provides larger transmission range than cars [8], and 3) governments do not need to pay an upfront infrastructure cost to deploy RSUs [8]. Two types of multi-hop are used in non-obstacle shadowing regions: 1) V2V when the vehicles are in transmission range of each other, and 2) V2I when the vehicles are not in the transmission range of each other. The multi-hop technique removes the long delays arising from the time required to deploy vehicular clouds.

The difference between Hybrid-Vehcloud and the previous approaches [3-7] is the consideration of obstacle shadowing region during message dissemination. Previous methods used multi-hop, roadside infrastructures or vehicular cloud to disseminate messages and hence, messages get dropped in the middle without reaching the destination in obstacle shadowing regions. whereas Hybrid-Vehcloud uses the vehicular cloud for disseminating messages in obstacle shadowing regions and multi-hop technique in non-obstacle shadowing regions to reduce message drops, ensure efficient resource utilization, rapid transmission of messages, reduced delay and better QoS.

Our *objective* is to lower the delay and to provide guaranteed message delivery obstacle shadowing regions having high vehicle densities. We considered three previous schemes for comparing with Hybrid-Vehcloud: 1) Cloud-assisted Message Downlink Dissemination Scheme (CMDS) [7], 2) Cross-Layer Broadcast Protocol (CLBP) [5], and 3) Cloud-VANET protocols [6]. These schemes were chosen as they used the vehicular cloud or multi-hop for disseminating messages, similar to Hybrid-Vehcloud.

*Results* from simulations using Simulation of Urban Mobility (SUMO) and network simulator (ns-2) show that Hybrid-Vehcloud provides guaranteed message delivery and reduced latency, and performs up to 30% better than previous techniques at high vehicle densities. The *contributions* of our work are:

1. We proposed and evaluated Hybrid-Vehcloud, a new technique to disseminate messages in both obstacle shadowing and non-obstacle shadowing regions.
2. Hybrid-Vehcloud yields a faster and efficient solution for disseminating messages between the vehicles.

The rest of the paper structured as follows: Description of the problem is illustrated in Section II. The proposed solution for message dissemination is presented in Section III. Performance evaluation of our approach illustrated in Section IV. The simulation results were presented in Section V, before concluding the paper in Section VI.

## II. PROBLEM DESCRIPTION

The radio transmissions are heavily affected by shadowing effects commonly known as obstacle shadowing [3,4]. Finding a solution for this problem plays an essential role in dense regions like a Manhattan and other downtown areas where buildings block radio propagation [4]. Assume vehicle V1 needs to disseminate messages to the nearby vehicles V2, V3, and V4. But, the shadow region, represented in Figure 1, blocks the radio wave propagation between vehicles.

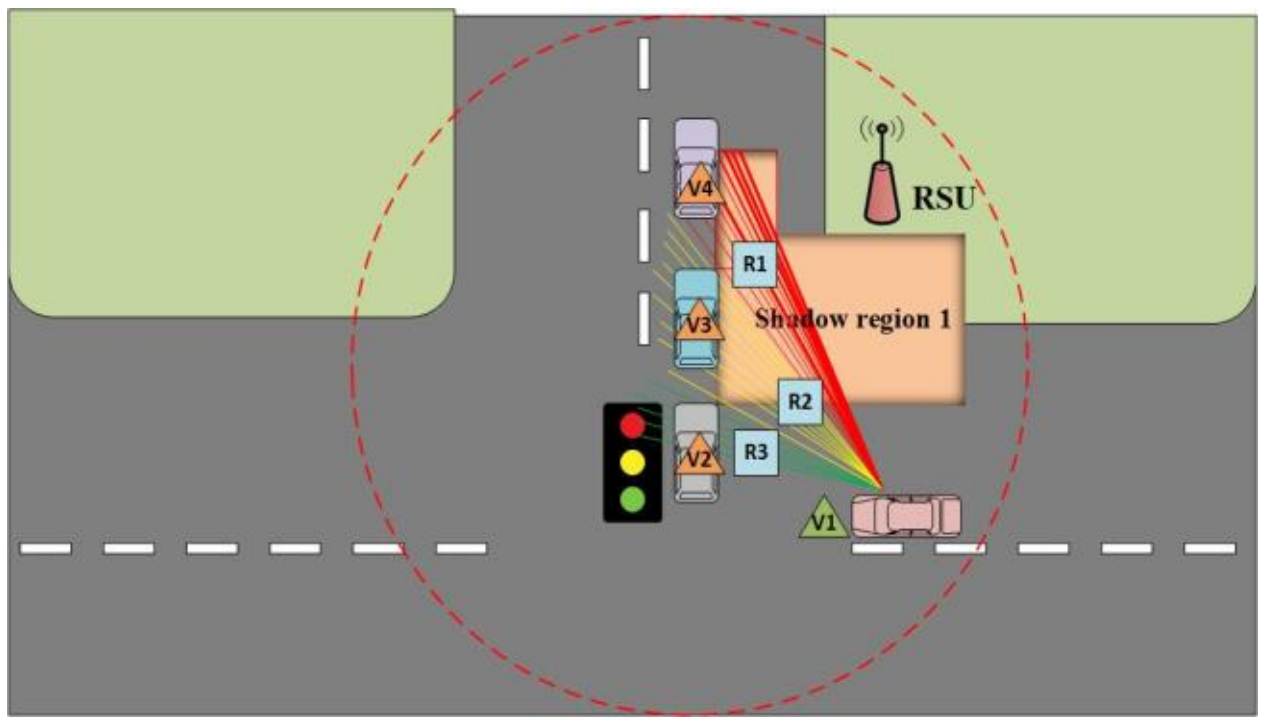

Figure 1: Obstacle shadowing caused by tall buildings, skyscrapers, etc. in dense vehicle regions.

To provide a solution, we divided this problem into three regions ($R_1$, $R_2$, and $R_3$) represented in the form of red, yellow and green lines respectively, where, $R_1$ is the obstacle shadowed region, $R_2$ is the partially shadowed or uncertain region, and $R_3$ is the non-obstacle shadowed region.

The vehicle V2 is in region $R_3$ (non-obstacle shadowed region), hence the messages can be disseminated using a hopping technique, whereas, vehicle V4 is in region $R_1$ (obstacle shadowed region) and its radio transmissions blocked due to shadowing effects. It leads to a situation where the messages dropped without reaching the destination. The vehicle V3 is in region $R_2$ (uncertain region) where the message may be sent directly or dropped without reaching the destination which increases the complexity of the system. To reduce the complexity and to increase the probability of message delivery we combined the regions $R_2$ and $R_1$ into a single region $R_1$.

To overcome the shadowing effects caused by tall buildings in a Manhattan and other downtown regions, we developed a hybrid technique called Hybrid-Vehcloud. It disseminates the message such as accident information, traffic congestion, potholes, etc. to nearby vehicles. The detailed explanation of Hybrid-Vehcloud is illustrated in Section III.

## III. PROPOSED SOLUTION

To provide a solution to the message dissemination problem in obstacle shadowing regions (Section II) and to disseminate messages in non-obstacle shadowing regions, we developed a technique called Hybrid-Vehcloud, represented in Figure 2. It consists of two cases: 1) dissemination of messages using the vehicular cloud (Case 1), and 2) dissemination of messages using multi-hop (Case 2).

**Case 1: Dissemination of messages using the vehicular cloud**

Assume the vehicles represented in Figure 2(a) are in obstacle shadowing regions and need to disseminate messages between each other. Though the vehicles are in communication range of an RSU, it is not possible to establish communication due to problems caused by shadowing. The solution for this situation is to deploy a vehicular cloud where the buses act as mobile gateways (Section I).

Following steps are involved in disseminating messages: First, vehicles collect information like the traffic jam, roadblock, etc. and disseminate them to mobile gateways using DSRC protocol. The mobile gateways deploy vehicular cloud based on infrastructure such as RSU and disseminate the message received from vehicles to the cloud. In addition, the mobile gateways transmit their own information like location, access delay, bandwidth etc. to the cloud. After receiving the information about gateways and input messages, the cloud servers assess the traffic density and determines suitable mobile gateways to disseminate the message. The gateways are selected to maximize the coverage range of vehicles in the targeted area. The messages are then transmitted to the vehicles using DSRC through an appropriate gateway. As mobile gateways are aware of the location of the vehicles probing the situation is not necessary to determine the obstacle shadowing regions.

**Case 2: Dissemination of messages using multi-hop**

Assume the vehicles represented in Figure 2(b) are in non-obstacle shadowing regions and need to disseminate messages between each other. Here, the messages are disseminated using multi-hop technique. Two types of multi-hop are used in non-obstacle shadowing regions: 1) V2V communication when the vehicles are in transmission range of each other, and 2) V2I communication when the vehicles are not in the transmission range of each other, where hopping takes between the vehicles and RSU. The multi-hop technique is adopted to disseminate the messages in non-obstacle shadowing regions due to the following reasons: 1) radio transmission of RSU in non-obstacle shadowing regions is unaffected due to the absence of obstacles like tall buildings, 2) to avoid large delays involved in deploying the vehicular cloud and disseminate messages between vehicles.

Based on the problem (Section II) and proposed solution (Case 1 and Case 2) discussed above, we formulated the Hybrid-Vehcloud as follows:

### 3.1. Obstacle modeling

Obstacle modeling is formulated based on our problem description (Section II) and proposed solution (Case 1 and Case 2). Assume the transmission range of a vehicle ($T_{\text{base}}$) is in the form of a circle and divided into two regions such as $R_1$, and $R_2$, where $R_1$ is the non-obstacle shadowed region, and $R_2$ is the

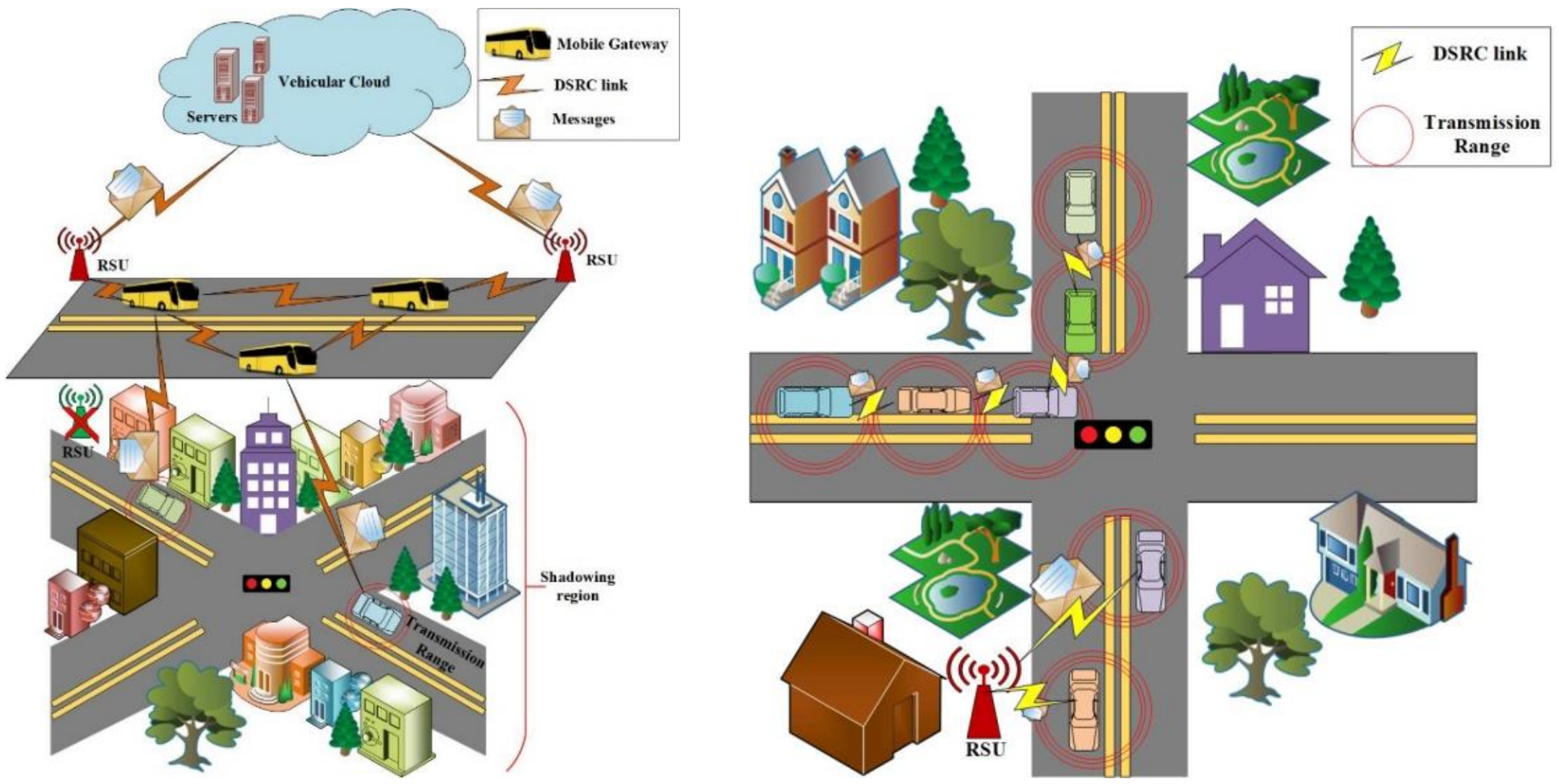

(a) (b)

Figure 2: Dissemination of message using Hybrid-Vehcloud**:** a) vehicular cloud technique and b) multi-hop.

-obstacle shadowed region ($O_{\text{shadow}}$). Usually, obstacle shadowed regions are represented in dB but, to calculate the area of regions R$_1$ and R$_2$ we need to express the $O_{\text{shadow}}$ in meters. It can be done by:

$$d = \frac{10(O_{\text{shadow}} - 32.44 - 20\log(f))}{20} * 1000 \quad (1)$$

Where $d$ is the distance, and $f$ is the frequency. For our approach, $f$ = 2.45 GHz.

The transmission range of a vehicle ($T_{\text{base}}$) and area of the regions associated in it (R$_1$ and R$_2$) are calculated as follows: 1) $T_{base} = \pi r^2$, 2) $R_1 = T_{\text{base}} - d$, and 3) $R_2 = d$.

For each obstacle in the line of sight between the vehicles, represented in Figure 1 and Figure 2(a), the effect of obstacle shadowing ($O_{\text{shadow}}$) region is calculated as follows:

$$O_{\text{shadow}} = \alpha n + \beta l_{\text{obs}} \quad (2)$$

Where $n$ is the number of times an obstacle encountered, $l_{\text{obs}}$ is the total length of an obstacle, $\alpha$ represents the attenuation due to the exterior wall, and $\beta$ represents the approximate internal structure of an obstacle. The generic equation to calculate the power received at a receiver end due to obstacle shadowing formulated as follows:

$$R_{\text{p}} = T_{\text{p}} + A_{\text{t}} + A_{\text{r}} - O_{\text{shadow}} \quad (3)$$

Where $R_{\text{p}}$ is the received power, $T_{\text{p}}$ is the transmitted power, $A_{\text{t}}$ is the antenna gain at the transmitter end, and $A_{\text{r}}$ is the antenna gain at the receiver end. Based on the transmitted power and received power in equation (3), we can determine the obstacle shadowing in regions like Manhattan environment and other downtown area at high vehicle densities.

### 3.2. Delay analysis

Delay refers to the time taken for a packet to be transmitted across a network from source to destination. It is an additive metric, and thus, overall delay (end-to-end delay) equal to the sum of delays in each hop during a multi-hop data transmission (Case 1 and Case 2) [9]. The single-hop delay of a network ($D$) can be calculated as follows:

$$D = t_{\text{trans}} + t_{\text{q}} + t_{\text{cont}} + t_{\text{proc}} + t_{\text{prop}} \quad (4)$$

Where $t_{\text{trans}}$ is the transmission delay, $t_{\text{q}}$ is the queuing delay, $t_{\text{cont}}$ is the contention delay, $t_{\text{proc}}$ is the processing delay, and $t_{\text{prop}}$ is the propagation delay.

### 3.3. Message success rate analysis

Message success rate directly impacts the performance of the system and thus, increase in message success rate improves the performance of Hybrid-Vehcloud. It is calculated as follows:

$$M_{\text{success}} = \frac{P_{\text{msg}} * D}{N_{\text{users}}} \quad (5)$$

$$\begin{cases} 0.5 \leq M_{\text{success}} \leq 1, & \text{message disseminated using multi hop} \\ 0 \leq M_{\text{success}} < 0.5, & \text{message disseminated using the vehicular cloud} \end{cases}$$

Where $M_{\text{sucess}}$ is the message success rate, $P_{\text{msg}}$ is the probability of message delivery, and $N_{\text{users}}$ is the number of users associated with the system.

From Equation (5), we can observe Hybrid-Vehcloud provides guaranteed message delivery to the nearby vehicles in an urban environment the vehicular cloud or multi-hop (Case 1 and Case 2), and thus, the robustness of our system is high when compared with previous techniques.

## IV. PERFORMANCE EVALUATION

The performance of Hybrid-Vehcloud is evaluated based on simulation using ns-2, and SUMO simulators. In this section, we discuss the simulation setup and various metrics involved in the simulation of Hybrid-Vehcloud.

### 4.1. Simulation setup

Simulation of Hybrid-Vehcloud is accomplished based on the techniques and scenarios discussed in Sections II and III, and the formulation of Hybrid-Vehcloud illustrated in Section III used in simulation to identify the obstacle shadowing regions in an urban environment (Section 3.1), and determines the technique to disseminate the message (Section 3.3). Moreover, it is used to calculate the delay associated with Hybrid-Vehcloud (Section 3.2). To simulate the trace of vehicles movements, we used open source traffic simulator SUMO. The output of the SUMO simulator given as input to the ns-2 simulator. NS-2 is a discrete event simulator consisting of many modules such as: 1) wireless signal propagation model for transmitting radio waves between the vehicles, 2) packet loss model to identify the number of packets dropped in transmission, 3) node deployment model for dynamic placement of nodes, 4) node mobility model for dynamic network topologies, etc. to perform the simulation. The simulations were performed based on the parameters, represented in Table 1.

Table 1: Parameters used in the simulation.

| Parameters | Value |
|---|---|
| Road length | 10 km |
| Number of vehicles/nodes | 50-450 |
| Number of lanes | 3 |
| Vehicle speed | 30-50 mph |
| Transmission range | 300m |
| Message size | 256bytes |
| Simulator used | ns-2, SUMO |
| Data rate | 2Mbit/s |
| Technique used | Multi-hop, Vehicular cloud |
| Protocol | IEEE802.11p |
| CW Min/Max | 31/1023 |

### 4.2. Performance Metrics

The simulations were performed based on the equations formulated in Section III. We considered following metrics to evaluate the performance of Hybrid-Vehcloud and to compare our results with CMDS, CLBP, and Cloud-VANET protocols (Section I):

- End-to-end delay: Time is taken for a message to be disseminated across a network from source to destination [9].
- Collision ratio: The number of packets colliding across a network before reaching the destination [1].
- Probability of message delivery: The probability of the input message delivered to the targeted vehicles [1].
- Average throughput: Average rate of successfully disseminated messages across a communication channel [10].

## V. SIMULATION RESULTS

As discussed in Section IV, the simulation of Hybrid-Vehcloud is performed based on the metrics such as: 1) end-to-end delay, 2) probability of message delivery, 3) collision ratio, and 4) average throughput using ns-2, and SUMO simulators. The results are discussed below:

**End-to-end delay:** In Hybrid-Vehcloud, knowledge of nearby vehicles including the position significantly reduces the route setup time and propagation time across a network. Hence, it delivers the message much faster when compared to CLBP, CMDS, and Cloud-VANET protocols, represented in Figure 3.

The end-to-end delay is calculated against the number of vehicles, and it increases as the number of users increases in the system due to a large number of messages need to be delivered within a specific time interval. It is formulated in Section 3.2.

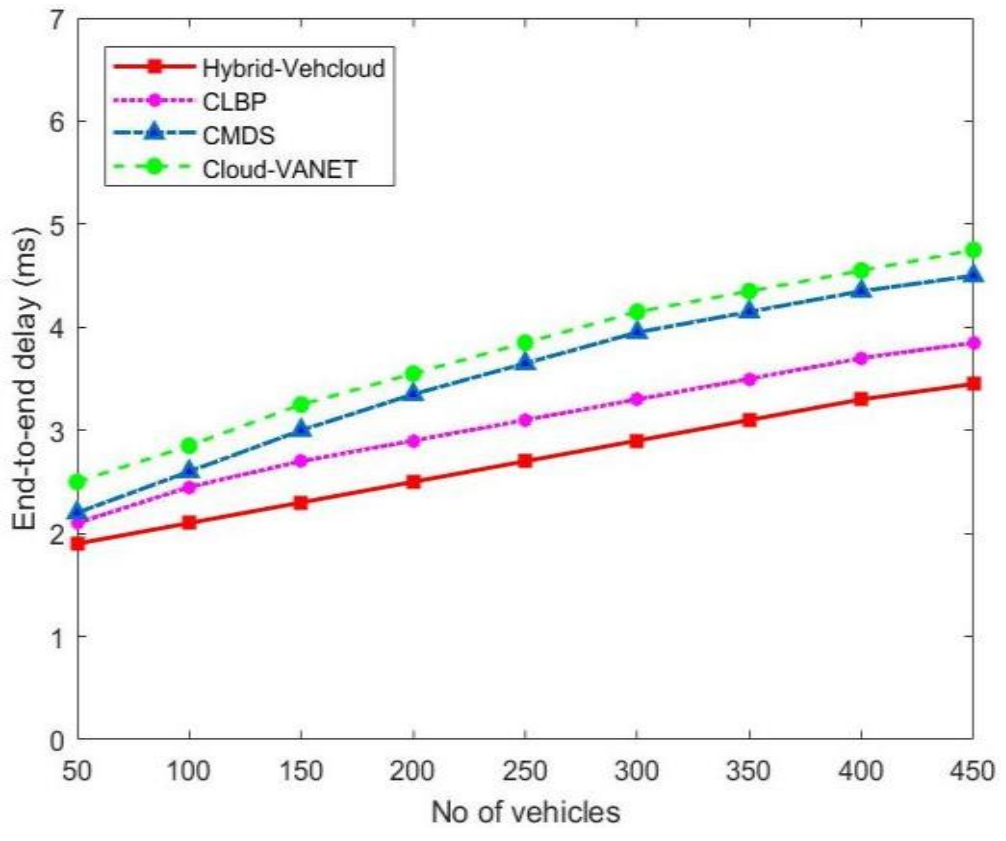

Figure 3: Comparison of end-to-end delay of Hybrid-Vehcloud with previous approaches.

**Probability of message delivery:** The probability of message delivery of Hybrid-VehCloud was observed to be higher due to guaranteed message delivery to the vehicles in obstacle shadowing region (Section III), represented in Figure 4. It is preformed based on the equation formulated in Sections 3.1 and 3.3.

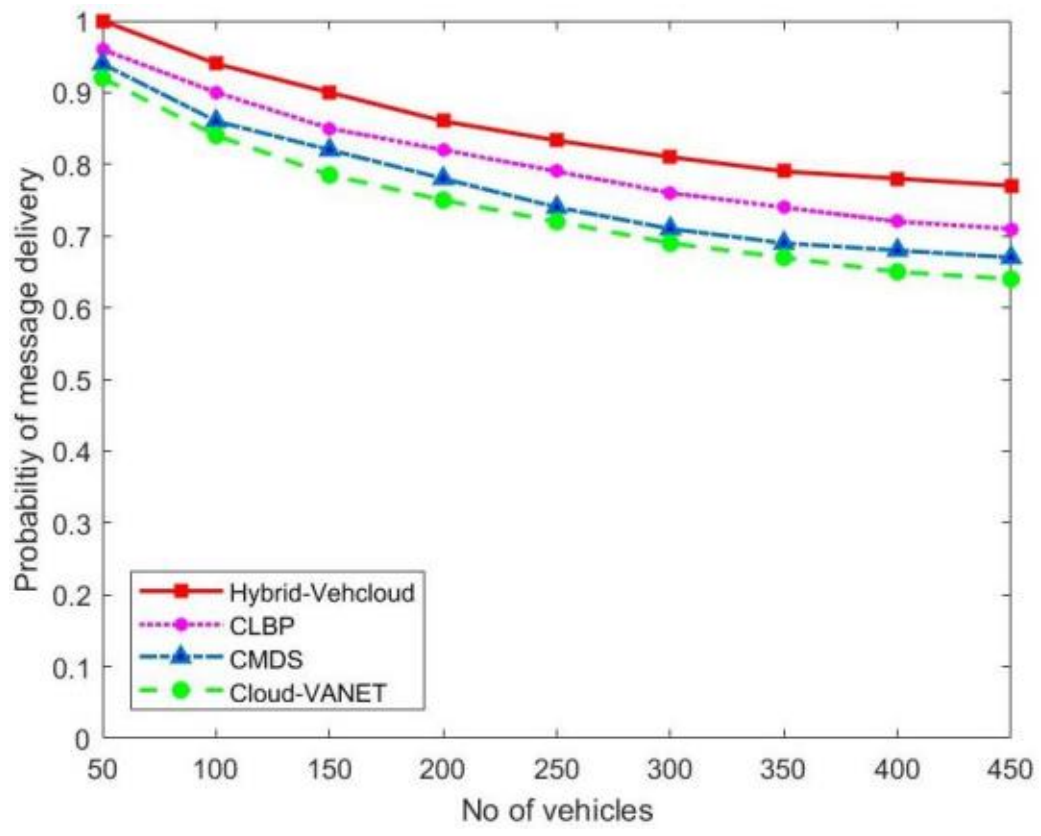

Figure 4: Comparison of probability of message delivery of Hybrid-Vehcloud with previous approaches.

We considered the probability of message delivery concerning the number of vehicles. For each user, the probability of message delivery distributed in the range of (0-1). From the Figure 4, we can observe that the probability of message delivery decreases marginally as the number of users increases due to the increase in load on Hybrid-Vehcloud.

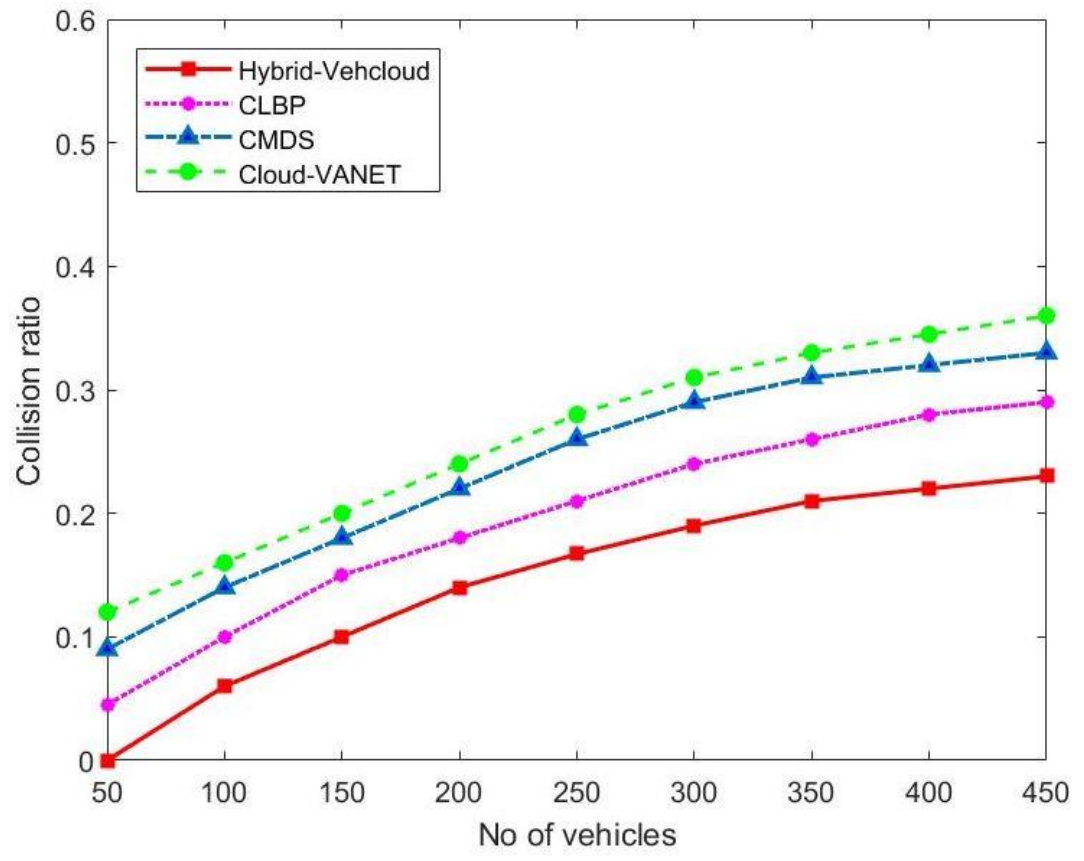

Figure 5: Comparison of collision ratio of Hybrid-Vehcloud with previous approaches.

**Collision ratio:** Collision ratio of Hybrid-Vehcloud is the number of packets colliding before reaching the destination. It observed to be lower at high vehicle densities and increases slightly due to the collision of packets as the number of users increases in a system, represented in Figure 5.

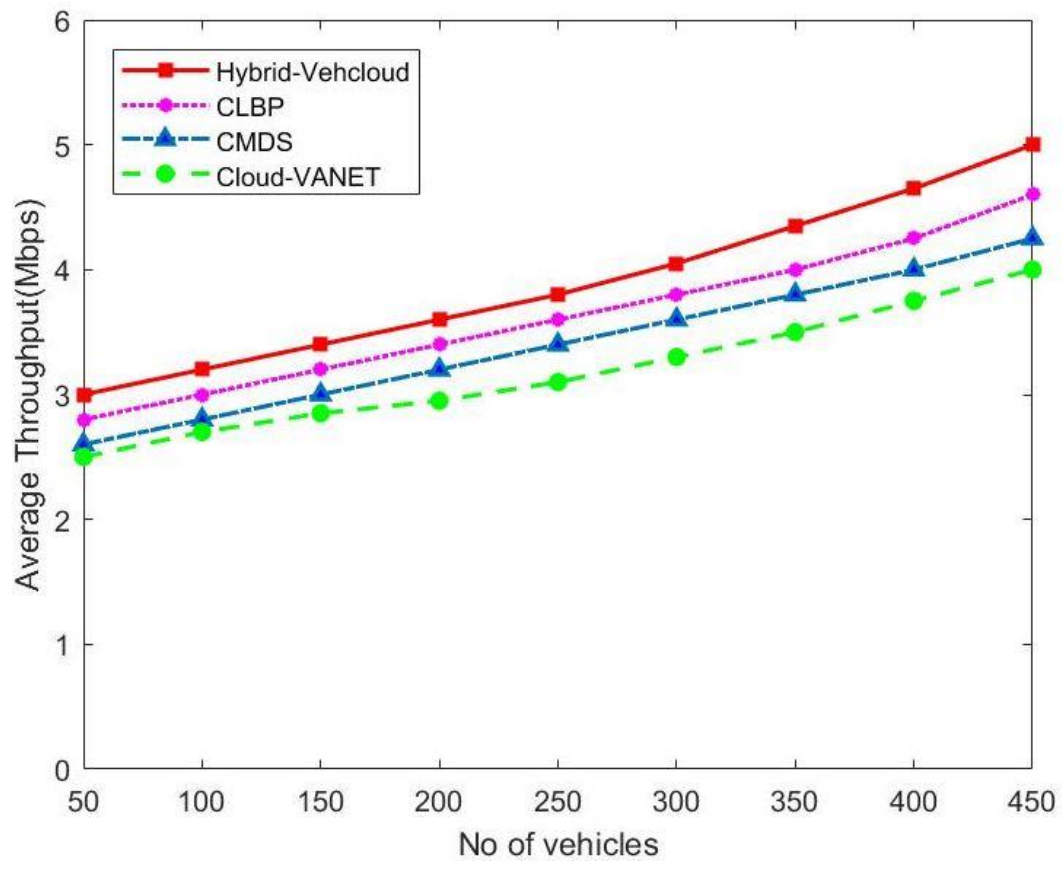

Figure 6: Comparison of average throughput of Hybrid-Vehcloud with previous approaches.

**Average throughput:** In Hybrid-Vehcloud, average throughput is the number of messages disseminated across a communication channel. From Figure 6, it can be observed that average throughput increases as the number of vehicle increases in the system, due to a large number of messages disseminated between vehicles. The average throughput of Hybrid-Vehcloud is high when compared to CLBP, CMDS, and Cloud-VANET protocols at all vehicle densities.

## VI. Conclusion

We addressed various vehicle disconnection problems in a vehicle-dense Manhattan like VANET environment in which transmitted messages may be lost due to obstacle shadowing. To address these problems, we developed a hybrid technique called Hybrid-Vehcloud which adopts mobile gateways (such as busses) in the vehicular cloud for messages dissemination in obstacle shadowing regions, and multi-hop technique for regions with no obstacle shadowing. It ensures lower delay and guaranteed message delivery. We have analyzed the performance of our proposed Hybrid-Vehcloud and performed extensive simulations using the ns-2 and SUMO simulators. The results showed that Hybrid-Vehcloud is robust, efficient, and provides the best performance when compared to CLBP, CMDS, and Cloud-VANET protocols at high vehicle densities.